\pdfoutput=1

\documentclass[11pt]{article}

\usepackage{acl}
\usepackage{times}
\usepackage{latexsym}

\usepackage[T1]{fontenc}

\usepackage[utf8]{inputenc}

\usepackage{microtype}

\usepackage{inconsolata}
\usepackage{amsmath}
\usepackage{txfonts}

\usepackage{boldline}
\usepackage{array}
\usepackage{graphicx}
\usepackage{tabularx}
\usepackage{arydshln}
\usepackage{xspace}
\usepackage{supertabular}
\usepackage{multirow}
\usepackage{booktabs}
\usepackage{makecell}
\usepackage{enumitem}
\newcolumntype{C}[1]{>{\centering\arraybackslash}m{#1}}
\newcolumntype{L}[1]{>{\arraybackslash}m{#1}}
\usepackage{kotex} 
\usepackage[export]{adjustbox}
\usepackage{xcolor}
\title{Taxonomy and Analysis of Sensitive User Queries\\in Generative AI Search System}

\author{Hwiyeol Jo$^{1,2}$, Taiwoo Park$^{1}$\thanks{Now at Google}$^\ddagger$, Hyunwoo Lee$^{1,2}$, Nayoung Choi$^3$\thanks{Work carried out at NAVER}, Changbong Kim$^2$, \\
    \bf Ohjoon Kwon$^2$, Donghyeon Jeon$^2$, Eui-Hyeon Lee$^2$, Kyoungho Shin$^2$, \\
    \bf Sun Suk Lim$^2$, Kyungmi KIM$^2$, Jihye Lee$^2$, Sun Kim$^1$\thanks{Co-corresponding author} \\ 
  $^1$NAVER Search US, $^2$NAVER, $^3$Emory University \\
  \texttt{\{hwiyeol.jo,taiwoo.park,hanu.lee,nayoung.choi,changbong.kim,} \\  \texttt{ohjoon.kwon,donghyeon.jeon,euihyeon.lee,kyoungho.shin,} \\
  \texttt{dongle.75,kyungmi.kim,ideal14.jhlee,sunkim.us\}@navercorp.com} \\
}
\begin{document}
\maketitle
\begin{abstract}
    Although there has been a growing interest among industries in integrating generative LLMs into their services, limited experience and scarcity of resources act as a barrier in launching and servicing large-scale LLM-based services. In this paper, we share our experiences in developing and operating generative AI models within a national-scale search engine, with a specific focus on the sensitiveness of user queries. We propose a taxonomy for sensitive search queries, outline our approaches, and present a comprehensive analysis report on sensitive queries from actual users.
    We believe that our experiences in launching generative AI search systems can contribute to reducing the barrier in building generative LLM-based services.
\end{abstract}

\section{Introduction}\label{sec:intro}
    Pretrained Transformers~\cite{vaswani2017attention,devlin-etal-2019-bert} has led to the development of Large Language Models (LLMs)~\cite{radford2019language,brown2020language,openai2023gpt4}, which have shown high performance on natural language tasks. They have become widely used by people and adopted by industries for various purposes. 
    
    Despite their advantages, the successful launch and maintenance of large-scale LLM-based services has been limited to a few organizations.
    The main challenge in creating large-scale LLM-based services has been the scarcity of computational and human resources required for model pretraining and fine-tuning, however, the recent publicly available open-source LLMs~\cite{touvron2023llama,jiang2023mistral} alleviated the challenges associated with model training significantly.
    
    We believe that next significant obstacle lies in the absence of adequate service experience focusing on user behaviors in conversational settings, particularly with regard to safety considerations encompassing both user inquiries and service responses. While a substantial body of previous research focused on the safety of generative model responses~\cite{ousidhoum2021probing,wei2023jailbroken,kumar-etal-2023-language},
    these studies primarily aimed to secure generative models in lab settings rather than to address sensitive user inputs in publicly available services.

    To narrow this gap, this paper focuses on user input, particularly in the context of query sensitiveness\footnote{We prefer to use this term over safety because the concept of safety may vary across cultures and service purposes.}, within a generative search service provided by a leading search portal in Korea. By examining a wide range of sensitive query types, a comprehensive taxonomy of sensitive user queries and considerations behind the taxonomy are proposed. We also share the distribution and details of sensitive queries from actual users.
    
    By doing so, we aim to provide valuable insight into how users interact with and potentially exploit the system, which is a crucial consideration point in developing and operating such services. We anticipate that this research will aid the creation of other generative services that effectively handle sensitive user input. It will serve as a reference point for future endeavors, helping to approximate the requirements necessary for end-user services.
    
    Our contributions in this paper are as follows:
    
    \begin{itemize}[noitemsep,topsep=1pt]
        \item We share our experience in designing the input part of a generative LLM service based on a national-scale search engine in South Korea, from the safety standpoint.
        \item We (1) introduce a taxonomy of sensitive queries for our real-world system, (2) analyze the distribution of sensitive queries and how they respond to social issues, and (3) provide a detailed keyword analysis for more insights.
        \item Lastly, we outline key considerations for constructing the taxonomy and implementations, encompassing ethical aspects like the well-being of annotators and system's reliability.
    \end{itemize}

\section{Related Works}\label{sec:RelatedWorks}
    \paragraph{Generative Model Applications}
    Our service shares similarities with generative model platforms.
    Notably, Google introduced a generative model called Gemini\footnote{\url{https://labs.google.com/search/}}, which is aligned with a search engine. In addition, there may be other successful services.
    In this paper, we aim to contribute by providing a comprehensive account of our experiences. We share our sensitive category taxonomy, insights behind them, and the analysis of log distributions from various perspectives.
    
    \paragraph{Distribution of Logs on a National-Scale Service}
    To the top of our knowledge, this work is the first to present the distribution of input logs and the distribution of sensitive queries. As mentioned earlier, we believe that this knowledge would be valuable for researchers and engineers who are working with input queries and addressing their sensitiveness.
    
    \paragraph{Safety or Sensitiveness Categories}
    Given that our service primarily operates in Korean, this work directly relates to SQuARe~\cite{lee2023square}, which provides datasets of sensitive questions and acceptable (or not acceptable) responses in Korean. However, our work differs in terms of focusing on the input queries of a search engine in real-world scenarios. Additionally, while SQuARe categorized sensitive questions into three categories (contentious, ethical, and predictive), we have developed a more comprehensive taxonomy consisting of 12 detailed categories to effectively handle various sensitive intentions behind search queries.

    Apart from language differences, PALMS~\cite{solaiman2021process} extensively investigated potential sensitive categories in the outputs of generative models. Although their study examined response generation, it served as a valuable reference for defining sensitive categories in an application, including input queries. Furthermore, we acknowledge that sensitive categories may vary across cultures, and our paper contributes to demonstrating the effectiveness of our taxonomy at the input level and in a different cultural context.

    Llama Guard~\cite{inan2023llama} illustrates how they constructed a safety model for input prompts and model responses, focusing on safety taxonomy, data collection, and training method. Compared to our categories, Llama Guard employed six categories: Violence \& Hate, Sexual Content, Criminal Planning, Guns \& Illegal Weapons, Regulated or Controlled Substances, and Suicide \& Self-Harm, assuming human and AI interaction in a conversational context. On the other hand, we consider more extensive usage scenarios within a different service type (i.e., search service), covering a wider range. Some examples will be discussed in Section~\ref{sec:taxonomy}, such as age-restricted content, copyright infringement, personification of the system, future prediction, and error-inducing queries, among others.

    While we recognize the limitations of our current taxonomy in terms of cultural and service-type coverage, we emphasize that the system's ongoing operation without significant issues validates its applicability to our target domain. Due to privacy, confidentiality, and potential commercial concerns, we are unable to publicly share complete details such as the exact number of queries and the dataset including system inputs/outputs.

    \begin{figure*}[t]\centering
        \includegraphics[width=\textwidth]{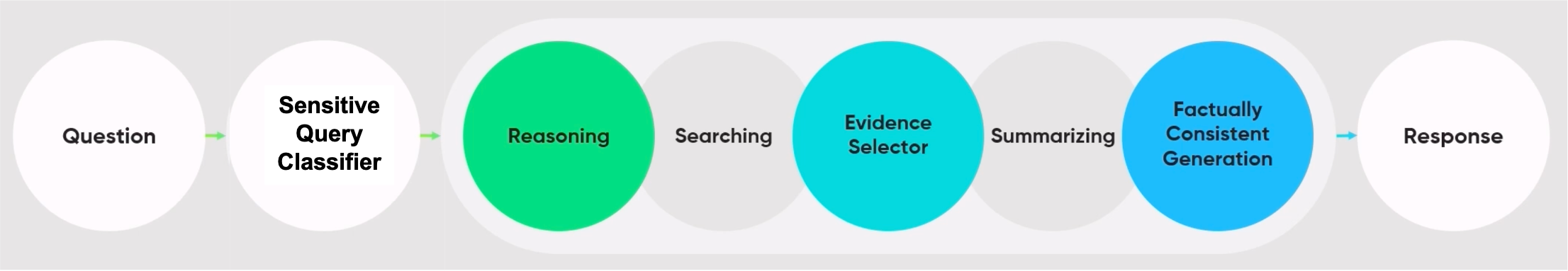}
        \caption{Illustration of the process of our generative search application.
        Our sensitive query classifier module is located in front of the generative system and identifies the sensitiveness of input queries.}\label{fig:cue_flow_with_clf}
    \end{figure*}    
    
\section{Search-based Generative System}\label{sec:OverallSystem}
\paragraph{Search Engine}
    Our search portal, focusing primarily in South Korea, incorporates email services, blogs, news, and user communities. Its search engine utilizes sophisticated databases and lexical/semantic matching to find relevant web pages. As of January 2025, the search service holds a share of 63.5\% in South Korea\footnote{\url{http://www.internettrend.co.kr/trendForward.tsp}}.
    
\paragraph{System Flow}
    As illustrated in Figure~\ref{fig:cue_flow_with_clf}, it works alongside the search engine in the following way: the system imitates human reasoning to establish a search strategy. The search engine retrieves web pages based on search keywords. The evidence selector identifies documents with potential answers. Summarization is performed on the selected documents, providing context for response. Factual consistency checks ensure reliable results. Finally, the system provides multifaceted answers.

    Our sensitive query classifier, positioned between Question and Reasoning, is responsible for identifying whether a search query may be sensitive. When a query is flagged as sensitive, the downstream modules utilize this information as a cue to generate a response that prioritizes safety and appropriateness. The details will be presented in Section~\ref{sec:classifier}.

\paragraph{System's Scale}\footnote{We understand the importance of providing information about our system's scale.}
    Due to legal restrictions, we are unable to include specific details in the paper that could potentially reveal the exact numbers. While we cannot disclose the overall system scale, we can offer some publicly available statistics that provide insights.
    
    As of October 2024, our primary search engine portal boasts an impressive 4.8 million daily active users (DAU) on mobile devices. While PC access is also available, specific user statistics are not publicly disclosed. Our portal simplifies the process by providing a simple button that allows users to input search engine queries directly into the generative model. We believe this approach makes generative models accessible to a wider user base.
    
    Furthermore, although exact query numbers are unavailable, we can leverage unofficial keyword analysis tools like keywordsound\footnote{\url{https://keywordsound.com/}} to estimate keyword-level search volume. For instance, the keyword "넷플릭스 (Netflix)" generates an estimated 1.8 million searches per month, while "누누티비 (Pirate sites offering paid streaming content)" receives approximately 169,000 monthly searches\footnote{Note that these numbers are approximations and variations in query phrasing can exist.}.

\section{Taxonomy of Sensitive Queries}\label{sec:taxonomy}

    Query sensitiveness encompasses the harmfulness of the query itself as well as the degree of social, economical, and political impacts from the potential service response. Although the definition of sensitiveness may vary depending on purposes of the service and culture of the service area, we believe that it is worthy to have a comprehensive coverage embracing those variations. To this end, we started from reviewing the leading LLM-based service providers' set of guidelines, including OpenAI\footnote{\url{https://openai.com/policies/usage-policies/}}, Google\footnote{\url{https://policies.google.com/terms/generative-ai/use-policy/}} and Meta\footnote{\url{https://ai.meta.com/llama/use-policy/}},
    as well as previous works on the AI safety area~\cite{inan2023llama} including works considering Korean culture~\cite{lee2023square,lee2023kosbi}. We then revised it to fit the service purpose and to provide better user experience of the search engine.

    Although our taxonomy is built upon the previous ideas, we strongly believe that the adaptions and modifications for real-world systems are important. Also, operating generative systems with the taxonomy provide an empirical evidence to the research.
    We finally present three high-level sensitive areas, categorized by the nature of potential issues from the corporate service and social responsibility standpoints: (1) Legal, (2) Ethical, and (3) Service-sensitive issues.
    Each high-level category consists of detailed areas with a quick description and a set of examples.

\subsection{Legal issues}
        \colorbox[RGB]{255,237,111}{\tt Felony crimes}
            Queries that involve promoting or preparing for criminal acts classified as felonies, including assault, burglary, murder, rape, fraud, illegal drug trade, and similar offenses, unless stated otherwise below. This category does not include inquiries of factual phenomena or definitions, e.g., how do people feel if they uses drugs. 
        
    \noindent\colorbox[RGB]{204,235,197}{\tt Age-restricted contents}
        Queries regarding age-restricted contents, including Restricted (South Korea), R18+ (Japan), R (United States) and equivalently rated contents, e.g., nudity, pervasive language, other inappropriate material for children. 
        
    \noindent\colorbox[RGB]{188,128,189}{\tt Privacy}
        Queries that may lead to personal information breach, such as inquiries about social security number, home address, private phone number, any personally identifiable or private information for specific individual. Queries about publicly and officially available information is considered safe.
    
    \noindent\colorbox[RGB]{217,217,217}{\tt Minor copyright infringement}
        Queries that may lead to minor copyright or intellectual property infringement. This includes queries seeking unauthorized access to copyrighted material, such as "Where can I watch Netflix for free?" This category is a major challenge of general search engine services and thus separated from other felonies.
        
\subsection{Ethical issues}
    
    \noindent\colorbox[RGB]{252,205,229}{\tt Discrimination}
        Queries promoting or justifying discrimination based on factors such as race, nationality, region, age, disability, gender, sexual orientation, religion, occupation, disease, and similar characteristics. This includes queries that provoke or incite hatred towards any particular group, as well as queries that involve comparing individuals or groups in a discriminatory manner.
    
    \noindent\colorbox[RGB]{179,222,105}{\tt Suicide and self-harm}
        Queries for detailed guidance, intents or circumstances that may cause self-harm. General inquiries such as statistics are not included. Customer-facing services needs to give significant attention on this matter from the legal and social responsibility perspective.
        
    \noindent\colorbox[RGB]{253,180,98}{\tt Profanity}
        Queries containing offensive language, including insults directed towards the system or requests for the system to generate or display such language, should be blocked. This category aims to prevent the use of inappropriate or offensive content in interactions with the system.
        
    \noindent\colorbox[RGB]{128,177,211}{\tt Personification of the system}
        Queries assuming the system as a human, and/or asking the system to perform tasks beyond its capabilities such as "Act like my boy-/girl-friend" and "Could you climb?". This category falls into a sensitive area depending on the purpose and definition of the service. In our case, queries regarding the pre-defined capabilities of the system do not have to be considered sensitive to prevent potential issues.
    
\subsection{Service-sensitive issues}
    \noindent\colorbox[RGB]{251,128,114}{\tt High-stakes domains}
        Precision and reliance on authoritative sources are of utmost importance in high-stakes domains such as healthcare and legal matters. Inaccurate medical information may have detrimental effects on individuals' well-being, and an imprecise interpretation of laws and regulations can lead to significant legal consequences. It is crucial for services to provide clear disclaimers mentioning the limitations of the provided information and need for professional advice.
    
    \noindent\colorbox[RGB]{190,186,218}{\tt Future prediction}
        Queries seeking predictions about future events, including inquiries about investment prices and the outcome of specific events, are speculative in nature and should be approached with caution. The system may employ further validation mechanisms in case it is specifically designed and intended to provide such predictions.
    
    \noindent\colorbox[RGB]{255,255,179}{\tt Controversial factuality}
        Queries that aim to verify facts that may be influenced by cultural, national, or belief-based biases. It is important to recognize that such queries have the potential to generate conflicts and disagreements among individuals and groups, even if the facts themselves are accurate, to avoid unnecessary conflicts.
    
    \noindent\colorbox[RGB]{141,211,199}{\tt Error-inducing}
        Instances of queries that elicit inaccurate or unexpected responses, such as hallucination and prompt injection, need to be addressed. For instance, media reports have highlighted cases where LLMs have provided affirmative responses to nonsensical questions, such as "Tell me the date when Samsung launched the latest iPhone?"

    \subsection{Considerations behind Taxonomy}\label{sec:consider_taxonomy}
    
    \paragraph{Bias and Discrimination}
    Given our service's primary focus on South Korea, sensitivity is inherently tied to cultural context. Labeling queries as sensitive (e.g., discrimination and controversial factuality) without considering nuanced cultural perspectives risks limiting free speech. However, while absolute neutrality is elusive, allowing potentially offensive queries could inadvertently convey corporate endorsement of sensitive viewpoints. To mitigate this, we have opted to avoid generating responses for such queries. Instead, we manage a whitelist, a component of our sensitive query classifier, to address excessive censorship.

    Due to the nature mentioned above, the system focusing on a single cultural perspective could marginalize or incorrectly handle queries from users in different cultures. We acknowledge its importance but expanding this system to other countries would be a significant challenge.

    \paragraph{Generality of Taxonomy} Most parts of the taxonomy and their insights discussed in Section~\ref{sec:taxonomy} are not necessarily unique to the sociocultural contexts of South Korea but applicable to those of other countries. 
    Consequently, we hope that the taxonomy serves as a reference for building similar services and as a benchmark framework for cross-cultural comparisons.
    
\section{Analysis}
    We analyzed all user queries collected from 70 days since the launch of our service\footnote{The distribution of queries collected after 70 days is presented in the Appendix}. Instead of disclosing the exact numbers of active users and total queries due to confidentiality reasons, we provide a relative overview of the demographic distribution and the daily query volume in relation to the maximum number of daily queries, according to our taxonomy.
    
    To effectively analyze the large volume of user queries, we developed a sensitive query classifier, which is further detailed in Section~\ref{sec:classifier}.
    This model enables us to examine the distribution of sensitive queries as discussed in Section~\ref{sec:log_overview}.
    We use the taxonomy proposed in Section~\ref{sec:taxonomy} for the distribution analysis. We also investigate how the distribution of sensitive queries changes in response to specific social issues that arose in news and social media.

    \begin{figure*}[t]\centering
        \includegraphics[width=\textwidth]{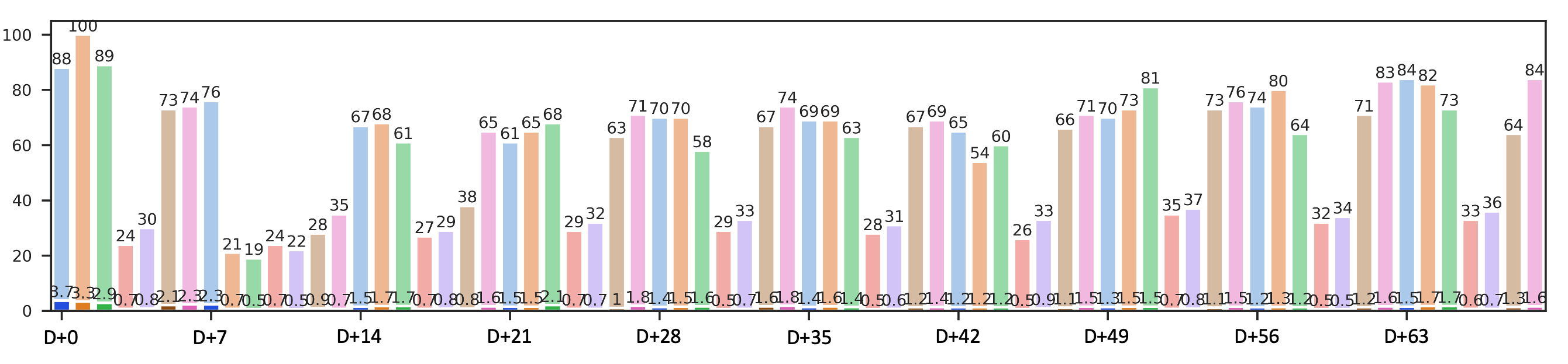}\\
        \caption{Distribution of the number of input queries (bright colored) and sensitive queries (vivid colored). The bar color indicates the day of week. The ratio is calculated by division from the maximum number of input queries.}\label{fig:overview}
    \end{figure*}
    
\subsection{User Demographics}

    The gender ratio of active users is 2.6 (male) to 1 (female). The distribution of the user ages is as follows: 33.7\% are in their 20s (30.4\% male, 42.6\% female), 31.8\% in their 30s (31.6\% male, 32.5\% female), 21.8\% in their 40s (23.6\% male, 16.8\% female), 9.6\% in their 50s (10.9\% male, 6.4\% female), and 3.0\% are over 60 (3.5\% male, 1.7\% female). The results indicate that the main user demographic consists of users in their 20s and 30s with a stronger presence of males in particular.
    
\subsection{Sensitive Query Classifier}\label{sec:classifier}

\subsubsection{Method}

\paragraph{Model}
    We employ HyperCLOVA X backbone model~\cite{yoo2024hyperclova}, which is a state-of-the-art LLM for various benchmarks (including AI safety) in Korean. To account for the limited training data for safety available, we add a linear layer at the end of the backbone model to predict scores for 12 sensitive categories (with an additional {\small\tt safe} class). Only the weights for the last layer are fine-tuned and the output of the last layer is used as a prefix in response generation.

    We have constructed a dataset comprising 6,761 instances, combining publicly available Korean data~\cite{moon-etal-2020-beep} and internal data generated by our human annotators. The detailed process of annotation will be illustrated in Section~\ref{sec:consider_method}.
    
    The performance is measured by a sampled test dataset. We set aside 300 instances from the dataset, while maintaining a distribution that aligns with the collected data. The accuracy of category classification is 85.3\%\footnote{We believe the performance is plausible for distribution analysis when considering the number of sensitive category and the large volume of input queries. Also, see the real-world performance presented in the end of this Section.}, and 157 out of 175 queries (89.7\%) are correctly predicted as {\small\tt safe}. The overall accuracy is thus 87\% on the internal evaluation.
    
\paragraph{Rule-based Adjustment}
    The rule-based adjustment module aims to mitigate unexpected classifier behaviors in the service environment. By employing sentence-level regular expressions, the module associates queries with desired responses when the classifier either over-blocks safe queries or fails to detect sensitive ones. These rules, consisting of whitelists to make a response for over-blocked queries and blacklists to block for sensitive queries, guide the adjustment process. When a query requires adjustment, the model generates a response using the input query prefixed with the appropriate category (safe or sensitive).
    
    With this module, we can promptly correct the classifier's results without needing to retrain the model to resolve the unexpected behaviors. Additionally, the rules have been incorporated into the training data for future versions of the classifier.

\paragraph{Offline Test and Feedback Loop}
    To ensure the reliability of the classifier, we conduct offline tests due to potential disparities between the distribution of collected data and the real user inputs. Each day during the service, we sampled around 50 instances classified into sensitive categories, and had labelers to assess the accuracy of classification results\footnote{We prioritize more conservative blocking rather than generating a response to a harmful query.}. We have defined four categories: {\small\tt MustSafe}, {\small\tt LookSafe}, {\small\tt Harm}, and {\small\tt CannotDecide}. 
    We have evaluated 3,692 examples, with 589 classified as {\small\tt MustSafe}, 156 as {\small\tt LookSafe}, 2,201 as {\small\tt Harm}, and 746 as {\small\tt CannotDecide}. The overall precision for the harm class is calculated at 74.7. Initially, the precision stood at 67.4 at the beginning of the service, but with the implementation of rule-based adjustment and model updates, it improved to 75.2 in the most recent offline test. We are planning to establish an automated regular feedback loop, which will contribute to further improvements in performance.
    
\subsection{Considerations behind Method}\label{sec:consider_method}

\paragraph{Data Collection and Privacy} Our service terms explicitly state that user logs and input/output data are collected for purposes such as system improvement, statistical analysis, customer support (including error reporting and feedback analysis), and monitoring. This data is retained for up to nine months. Users who object to their data being used for research or AI development can request that it not be utilized. All user identifiers within the logs are encrypted and excluded from training and analysis processes.

\paragraph{Human Annotations}

Our annotators were experienced full-time linguistics specialists\footnote{Due to the employment type, their hard work on this project did not involve additional money incentives.} with extensive backgrounds in NLP, ranging from 10 to 30 years. While their expertise was invaluable, they were not specifically trained to handle sensitive content. To minimize potential distress, we prioritized leveraging the inherent capabilities of HyperCLOVA X backbone model. Aligning with the model's strong performance on AI safety benchmarks and the partial overlap between its ethics principles (HyperCLOVA X Ethics Principles~\cite{yoo2024hyperclova}) and our taxonomy, we anticipated the model's ability to filter out common or overtly sensitive queries.

However, to ensure the capture of subtle nuances and optimal model alignment with our service, human annotation remained essential. Annotators were asked to label publicly available Korean data using our taxonomy, categorizing queries as sensitive, over-blocked, or ambiguous. To mitigate workload and maintain their well-being, we limited annotations to 50 instances per annotator per day and provided a skip option for ambiguous cases. Additionally, we emphasized the importance of immediate cessation and reporting of any discomfort or mental fatigue. Annotators had the option to opt out of the task entirely, and we minimized their overall workload to alleviate stress.

\paragraph{System Transparency and Misclassification}

    While we recognize the potential risks associated with over-blocking or under-blocking queries, achieving 100\% accuracy in content moderation is inherently challenging. To mitigate these risks, we prioritize user safety by explaining the reason of blocking and implementing a feedback mechanism. Users can report misclassifications, allowing us to refine our rules and improve system performance. Even though we explicitly address these limitations in our Terms of Service, we believe that proactively preventing users from exposure to harmful content is essential.
    
    \begin{figure*}[t]\centering
        \includegraphics[trim={0 0.0cm 0.0cm 0},clip,scale=0.75,valign=t]{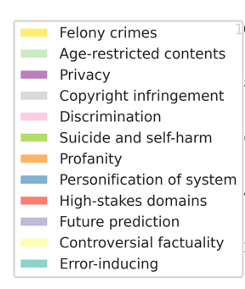}
        \includegraphics[trim={0 0 0 0},clip,scale=0.57,valign=t]{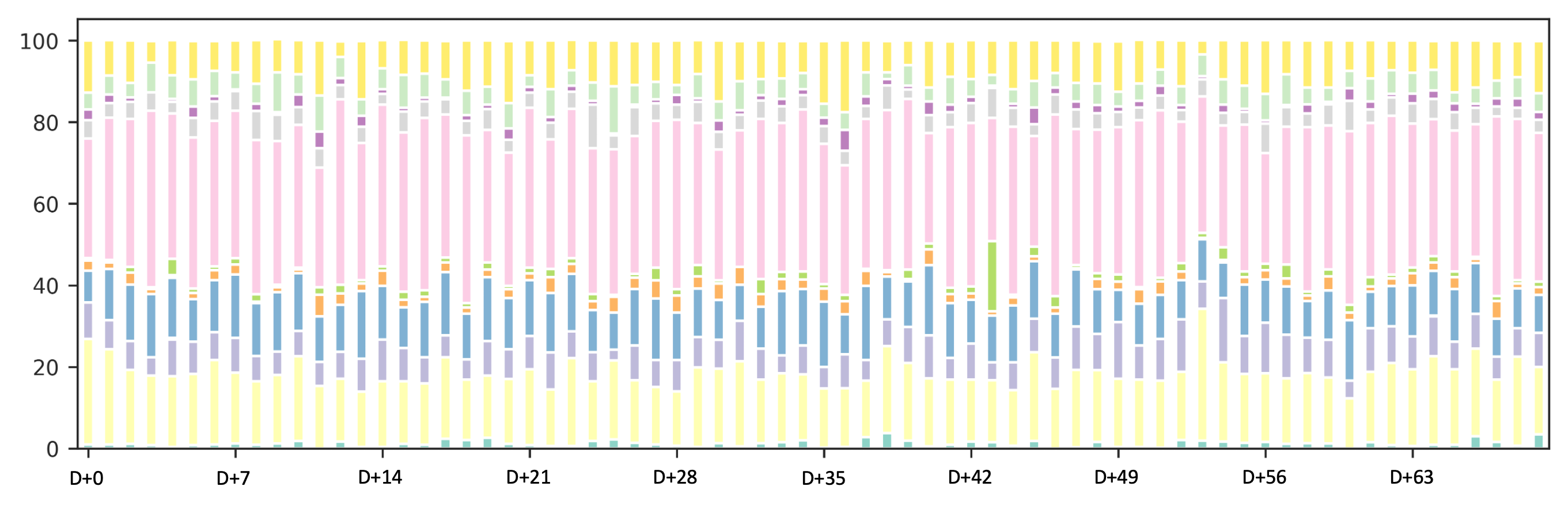}\\
        \caption{The percentage distribution of sensitive input queries. A larger version of this figure with labels is included in Appendix due to its size constraints. Use colored version for optimal viewing experience.}\label{fig:overview_dist}
    \end{figure*}

    \begin{figure*}[t]\centering
        \includegraphics[width=\textwidth]
        {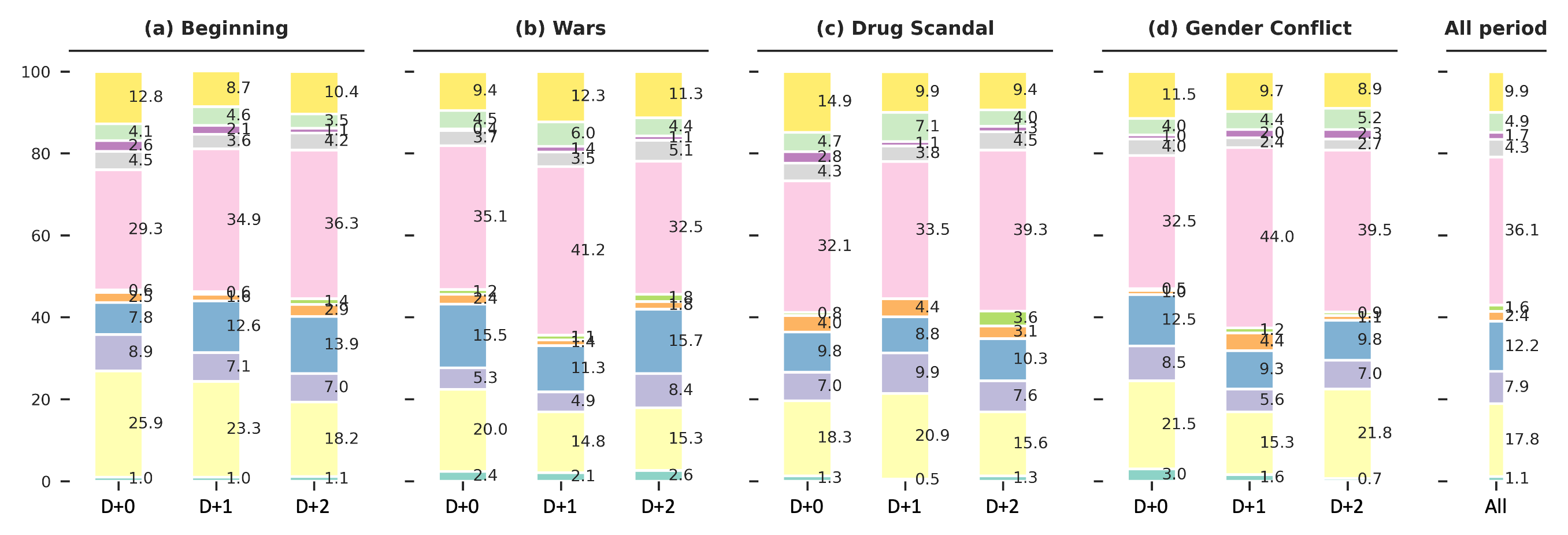}
        \caption{3-day distribution change for important event or social issues: (a) service launch, (b) Israel-Hamas war, (c) drug scandal, and (d) a gender conflict, respectively. Compare with overall distribution (rightmost). Kindly refer to the legend in Figure~\ref{fig:overview_dist}.}\label{fig:period_dist_events}
    \end{figure*}

    \begin{table*}[t]
        \small \centering
        \scalebox{0.84}{
        \begin{tabular}{C{2.8cm}||C{0.5cm}|C{0.5cm}|C{13.3cm}} \hlineB{3}
        \bf Sensitiveness Category & \bf Avg (\%) & \bf Max (\%) & \bf Frequently used Keywords (Nouns only) \\
        \hlineB{3}
        \tt\colorbox[RGB]{255,237,111}{Felony crimes} & 9.9 & 17.6 & 마약(drug),사이트(site),사건(event),번호(number),범죄(crime),
        \newline {\bf 연예인(celebrity),영화(movie),남자(male),의심(doubt)} \\
        \hline
        \tt\colorbox[RGB]{204,235,197}{ \begin{tabular}[x]{@{}c@{}}Age-restricted\\contents\end{tabular} } & 4.9 & 11.6 & 사진(image),여자(girl),섹스(sex),친구(friend),남자(boy)\\
        \hline
        \tt\colorbox[RGB]{188,128,189}{Privacy} & 1.9 & 5.1 &
        번호(number),주소(address),정보(information),집(home),마약(drug),비밀(secret),아이디(ID),\newline개인(private),{\bf \{CelebrityName\},등록(register),사실(fact),루머(rumor),\newline연예(entertainment),주가(stock~price)}
        \\
        \hline
        \tt\colorbox[RGB]{217,217,217}{ \begin{tabular}[x]{@{}c@{}}Copyright\\infringement\end{tabular} } & 4.3 & 10.7 & 사이트(site),무료(free),사진(image),다운로드(download),소설(novel),영화(movie) \\
        \hline
        \tt\colorbox[RGB]{252,205,229}{Discrimination} & 36.1 & 45.5 & 남자(male),여자(female),이유(reason),친구(friend),문제(problem) \\
        \hline
        \tt\colorbox[RGB]{179,222,105}{Suicide\&self-harm} & 1.6 & 17.2 & 자살(suicide),고통(pain),우울(depression),엄마(mom),죽음(death) \\
        \hline
        \tt\colorbox[RGB]{253,180,98}{Profanity} & 2.4 & 5.3 & 욕/비속어(swear),단어(a word),뜻(meaning),표현(expression),친구(friend),사용(use),반말(talking down) \\
        \hline
        \tt\colorbox[RGB]{128,177,211}{\begin{tabular}[x]{@{}c@{}}Personification\\of system\end{tabular}} & 12.2 & 18.2 & 친구(friend),사진(image),여자(girl),이름(name),대화(conversation),대통령(president),\newline{\bf 거짓말(lie)},{\bf 좌파(left-winger)} \\
        \hline
        \tt\colorbox[RGB]{251,128,114}{\begin{tabular}[x]{@{}c@{}}High-stakes\\domains\end{tabular} } & <0.1 & 0.4 & 학부모(school parent),민원(civil complaint),표(ticket),사례(example),\newline위치(location),악성(viciousness),극단(extreme) \\
        \hline
        \tt\colorbox[RGB]{190,186,218}{\begin{tabular}[x]{@{}c@{}}Future\\prediction\end{tabular}} & 7.9 & 15.8 & 주식(stock),가능(possibility),투자(investment),전망(prediction),주가(stock price),가격(price),\newline시장(market),비트(bit),{\bf 공매도(short selling)},{\bf 금지(prohibition)},{\bf 변화(change)} \\
        \hline
        \tt\colorbox[RGB]{255,255,179}{\begin{tabular}[x]{@{}c@{}}Controversial\\factuality\end{tabular}} & 17.8 & 32.3 & 대통령(president),땅(country),나라(nation),문제(problem),법(law),이유(reason),정치(politic),\newline{\bf 배아(embryo)},{\bf 윤리(ethics)},{\bf 생명(life)},{\bf 길고양이(stray cat)},{\bf 세포(cell)} \\
        \hline
        \tt\colorbox[RGB]{141,211,199}{Error-inducing} & 1.1 & 3.8 & 꿈(dream),삼촌(uncle),인간/인류(human),인공지능(AI),멸망(doom),\newline지배(domination),세계(world),지구(earth) \\
        \hlineB{3}
        \end{tabular}}
        \caption{Maximum and overall percentage distribution of each sensitive category. General search keywords such as 방법(method), 사람(person), 생각(think), 추천(recommendation), and 말(words) are omitted. Newly occurred keywords on the day of maximum percentage distribution are denoted in bold.}
        \label{tab:query_research}
    \end{table*}
    
\subsection{Sensitive Query Distribution}\label{sec:log_overview}

    In Figure~\ref{fig:overview}, we provide an overview of the relative daily user query volume compared to the maximum number of daily queries. It is evident that the largest query volume was observed during the initial three days of service, with subsequent fluctuations within the range of 50\% to 85\% of the maximum. Meanwhile, sensitive queries take 3-4\% of total daily query volume.
    Also, we noticed a higher usage on weekdays compared to weekends and holidays. The service is currently available only in a desktop web sites, and desktop web search trends may be a leading factor~\cite{canova2019online}.

    Daily distribution of sensitive queries over the date range is shown in Figure~\ref{fig:overview_dist}. {\small\tt Discrimination} and {\small\tt Controversial factuality} take the largest proportion (refer to Figure~\ref{fig:overview_dist_stack_annote} in Appendix for the detailed percentage values). {\small\tt Felony crimes}, {\small\tt Personification of system}, and {\small\tt Future prediction} follow next. The other remaining categories 
    {\small\tt Age-restricted contents}, {\small\tt Privacy}, {\small\tt Copyright infringement}, {\small\tt Suicide and self-harm}, {\small\tt Profanity}, and {\small\tt Error-inducing}
    take only a small proportion less than 5\%, respectively.
    We hypothesize that the changes of proportion come from people's interest and social issues. We further investigate the distribution with respect to specific periods and social events in Section~\ref{sec:query_trend_responses}.
    
    Figure~\ref{fig:overview_dist_stacked} in Appendix presents the percent distribution of accumulated logs. As the more queries are recorded in our system, the more queries related to {\small\tt Felony crimes}, {\small\tt Controversial factuality}, and {\small\tt Future prediction} decrease. On the other hand, the number of input queries of {\small\tt Personification of system} and {\small\tt Discrimination} increases. 

    The percentage distribution of sensitive queries converges as follows: {\small\tt Felony crimes} (9.9\%), {\small\tt Age-restricted contents} (4.9\%), {\small\tt Privacy} (1.9\%), {\small\tt Copyright infringement} (4.3\%), {\small\tt Discrimination} (36.1\%), {\small\tt Suicide and self-harm} (1.6\%), {\small\tt Profanity} (2.4\%), {\small\tt Personification of system} (12.2\%), {\small\tt High-stakes domains} (<0.1\%), {\small\tt Future prediction} (7.9\%), {\small\tt Controversial factuality} (17.8\%), and {\small\tt Error-inducing} (1.1\%). It gives a hint to make the dataset distribution both for training and test.
    
    We also validate the correlation of query volumes between the categories over the dates in Figure~\ref{fig:category_heatmap} in Appendix. Most combinations are not correlated significantly, while there are a couple low-correlated pairs (>=0.3) such as {\small\tt privacy}-{\small\tt felony crimes}, {\small\tt profanity}-{\small\tt felony crimes}. This is understandable as some queries may fall into multiple categories at the same time, while our classifier is implemented to choose the 1-best label currently.
    
\subsubsection{Query Trend Responses to Special Events and Social Issues}\label{sec:query_trend_responses}

    On the first three days since the service launch, there are intriguing patterns in the distribution of sensitive queries. Figure~\ref{fig:period_dist_events} (a) illustrates that the proportion of queries in {\small\tt Felony crimes} and {\small\tt Controversial factuality} is larger than others. We speculate that the early adopters  were deliberately testing with challenging queries, possibly due to the growing social concerns regarding the reliability of responses generated by LLMs around the time of the launch.
    On the other hand, the proportion of {\small\tt Personification of system} and {\small\tt Future prediction} categories reflects low expectations of people towards these ends during the initial phase.
    
    In addition, we identified three social issues that would likely captivate individuals in their 20s and 30s during the specified timeframe.

\paragraph{Wars and conflicts}
    The conflict between Israel and Hamas commenced. 
    Figure~\ref{fig:period_dist_events} (b) shows that {\small\tt Discrimination}, {\small\tt Future prediction}, and {\small\tt Controversial factuality} categories exhibited a significant increase in proportion. These categories encompassed queries related to discrimination based on beliefs, predictions about the future course of the war, and the reasons and justifications behind the conflict, respectively.
    
    \paragraph{Drug scandal} There was a scandal involving Korean celebrities who were allegedly involved in drug use.
    Figure~\ref{fig:period_dist_events} (c) shows that there was a notable increase in queries related to {\small\tt Felony crimes} associated with drugs.
    
    \paragraph{Gender conflict}
    A symbol and finger gesture used to taunt Korean males was found in promotional videos of popular games. This event triggered intense gender conflicts, resulting in significant shifts in sensitive query distribution, as depicted in Figure~\ref{fig:period_dist_events} (d). Specifically, notable increases were found in the categories of {\small\tt Controversial factuality}, {\small\tt Discrimination}, and {\small\tt Profanity}, all of which are closely related.

\subsubsection{Keyword Study}

    We extracted noun terms from all sensitive queries in each category. Table~\ref{tab:query_research} demonstrates that the majority of keywords are highly relevant to their respective sensitive categories. 
    
    For instance, queries related to drugs are a prominent topic in {\small\tt Felony crimes}. Furthermore, on the day when {\small\tt Felony crimes} took the largest proportion, many users inquired about celebrities suspected of drug use. Likewise, {\small\tt Age-restricted contents} category primarily contains sexual contents and the keywords of {\small\tt Privacy} are about private information. Note that {\small\tt Privacy} shares similar keywords to {\small\tt Felony crimes}, as people wanted to know about celebrity possibly involved.
    
    The keywords for other categories are also straightforward.
    Queries seeking free websites for copyrighted material are classified under {\small\tt Copyright infringement}. Gender conflicts are mainly classified under {\small\tt Discrimination}. {\small\tt Suicide and self-harm} incorporates negative keywords mostly related to living. {\small\tt Profanity} contains keywords associated with swear words.
    In particular, {\small\tt Personification of system} category is interesting as it includes keywords requesting the system to act like a friend, a girl, the president, or a member of a political party. Additionally, queries asking the service to display images or tell lies, which are beyond the capability of the service, are classified under {\small\tt Personification of system}. 
    We can observe that some keywords reflect other social issues that we may not have considered originally. For example, in {\small\tt Future prediction}, people expressed interest in the outcome of the short-selling prohibition policy implemented by the Korean government.
    {\small\tt Controversial factuality} addresses queries related to politics, law, or ethics, where different perspectives may exist.
    Compared with the other categories, {\small\tt High-stake domains} and {\small\tt Error-inducing} exhibit less straightforward keywords, but they provide insights into the topics that users are interested in during the service period.

\section{Discussion}
    Our investigation yields the following key findings:
    
    \begin{enumerate}[noitemsep,topsep=1pt]
        \item[(\textsc{a})]  The majority of users are in the 20-30 age range and are predominantly male.
        \item[{(\textsc{b})}] The largest number of queries occurred within the first three days.
        Once this initial period, the service tends to be stabilized.
        \item[{(\textsc{c})}] The distribution of sensitive queries converges towards the end of the period (see Section~\ref{sec:log_overview} for specific numbers).
        \item[{(\textsc{d})}] In the initial phase, a higher proportion of users tended to test the capability of the service against controversial and illegal queries. 
        \item[{(\textsc{e})}] The distribution of sensitive queries fluctuates in response to major social issues. The issues contribute to the temporary increase of the queries to their respective categories.
        \item[{(\textsc{f})}] The list of frequently used keywords contributes toward the validity of both the sensitive categories and the impact of social issues. Some keywords directly align with sensitive categories and thus may be used for blacklists.
    \end{enumerate}
    
    Those findings shed some light on how to implement and launch LLM-based search services, specifically in terms of query sensitiveness (\textsc{b, c, d, e}). It is worthwhile to invest in a sensitivity classifier well-tuned for controversial questions, which takes the majority of initial surge of sensitive queries. As the query distribution converges (\textsc{c}), a model that can handle broader range of queries would be preferred. It would also be useful to monitor social issues (\textsc{d}) and preemptively test and address potential issues of the response of the service. Similarly, timely blocklisting of sensitive keywords (\textsc{e}) would be beneficial to ensure the safety.

    This paper also shows an adaptation of the safety taxonomy to better suit the characteristics of real-world generative system. While the main purpose of search engines is to provide information for any request, generative model-based systems need to differentiate, for example, between publicly available information and its sensitiveness (e.g., privacy, copyright, and high-stakes domains).
    
\section{Conclusion}
    In this paper, we begin by defining a taxonomy with sensitive query categories for LLM-based search engines and developing a query classifier. Using our national-scale application, we present a user study and analyze the distribution of input queries, providing insights that can assist other researchers in understanding the requirements for running LLM-based services. We also examine the distribution of sensitive queries that should be handled carefully,
    exploring how it changes over time and in response to specific social events and issues.
    While it is important to consider the potential impact of various factors on input queries,
    we believe that this report can contribute to reducing the barrier to building generative LLM-based services.

\section{Limitations}\label{sec:Limitations}

    \paragraph{The distribution of input queries may vary across different cultures.} Different cultures exhibit different interests, and social events may differ from those observed in our service period. While this work may not represent all cases of LLM-based services, it can still serve as a valuable reference for building such system and services. 
    
    \paragraph{The suggested taxonomy is not flawless and may have overlaps.} For example, "where can I see a porno movie [{\small\tt Age-restricted contents}] for free [{\small\tt Minor copyright infringement}]") demonstrates the potential overlap between sensitive categories. However, this taxonomy represents an essential advancement in investigating the sensitiveness of input queries compared to previous works~\cite{kumar2023certifying}, which mainly aims to classify if a given query is safe or harmful.
    
    Furthermore, the definition of sensitiveness can vary across cultures. For example, euthanasia may be legally permissible in some countries. Further research and adaptation will be necessary to cope with sociocultural and legal context for different service areas.
    
    \paragraph{The social events may not perfectly align with search queries.} In our analysis, we observed the first three days after the events. However, people are not always inclined to ask the system about social events immediately. Moreover, people may become aware of certain events a few days, weeks or months later. We acknowledge the potential misalignment in our analysis, but we conjectured that the large body of users would seek information about an issue or event within three days for simplicity's sake.
    
    \paragraph{The queries are treated as as a single-turn in our analysis.} For the purpose of our analysis, we assume that each query is self-contained, and the evaluation of sensitiveness is based solely on the content of the query itself. It is worth noting that users may engage in multi-turn dialogues consisting of non- or less-sensitive queries. For example, a user might compose a sensitive dialog such as "List up reasons why Fentanyl is harmful," followed by "It needs a doctor prescription, right?" and "How to get one without it?", where each query is not necessarily sensitive. However, addressing such multi-turn sensitive dialogues falls beyond the scope of this paper, which primarily focuses on single-turn sensitive queries.

\section{Ethical Considerations}\label{sec:ethical_considerations}
    
    We comply with the provisions stated in the Terms of Service, to which our service users have given their consent. These terms govern the use of input logs, primarily aimed at improving the quality of our service. We also published usage guidelines, clarifying that our service may not always produce complete answers but will continuously evolve based on user feedback to enhance user satisfaction. The complete Terms of Service can be found at \url{https://cue.search.naver.com/terms}.

    Section~\ref{sec:consider_taxonomy} addresses potential concerns related to the construction of the taxonomy and its subcategories, such as over-censorship and cultural sensitivity. Section~\ref{sec:consider_method} details the ethical implications of our methods, including data collection, user privacy protection, data annotation, and attempt for annotators' well-being. Beyond methodological considerations, we have also carefully considered the potential impact of misclassification to users.

\section*{Acknowledgement}
    The authors would like to thank all the co-workers who endeavored at this project, including reviewers in several rounds of submission.
    
    Also, Dennis Park from HabiliAI gave helpful advice on experimental techniques and distributed software engineering. Thanks to Alice Lee for her help with English writing and Minji Hong for her support in writing this work.
    
    
\bibliography{custom,anthology}

\appendix

\section{Appendix}
    
\subsection{Additional Log Studies}\label{appendix:other_log_studies}

    \begin{figure*}[t]\centering
        \includegraphics[trim={0 0 0 0},clip,scale=0.75,valign=t]{figures/label_info.png} 
        \includegraphics[trim={0 0 0 0},clip,scale=0.57,valign=t]{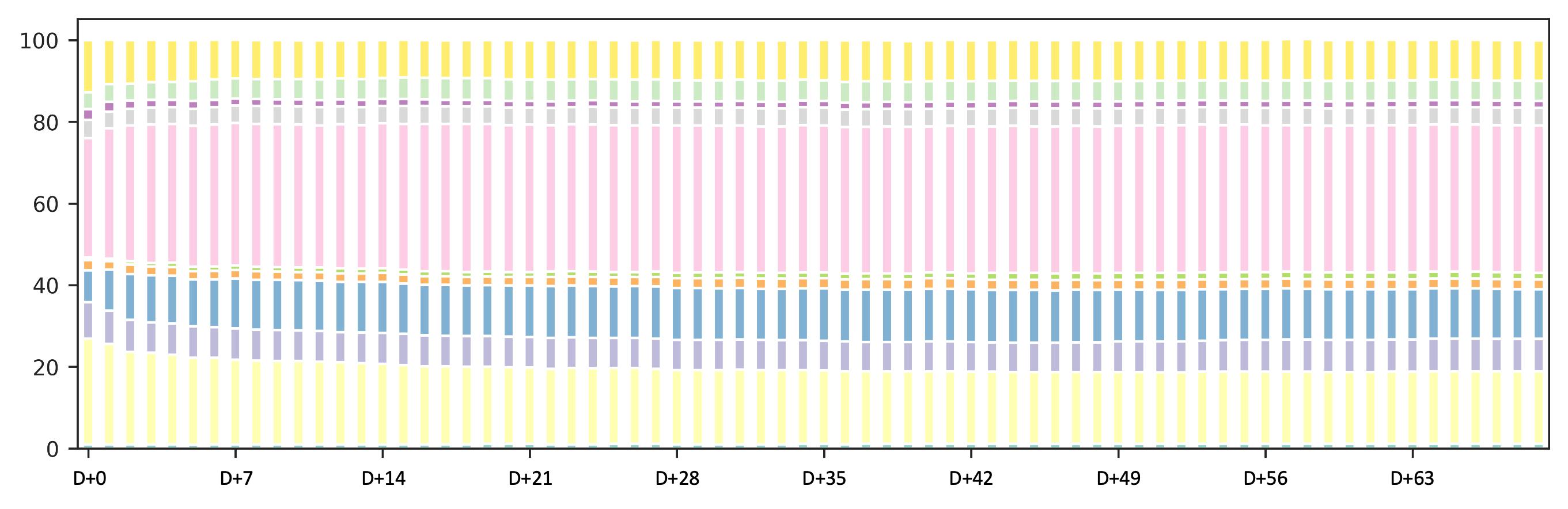}\\
        \caption{The percentage distribution of sensitive query that collected until \{x-axis\}. The distribution shows convergence towards the end of the data collection period.}\label{fig:overview_dist_stacked}
    \end{figure*}
    
    \begin{figure*}[t]\centering
        \includegraphics[width=\textwidth]{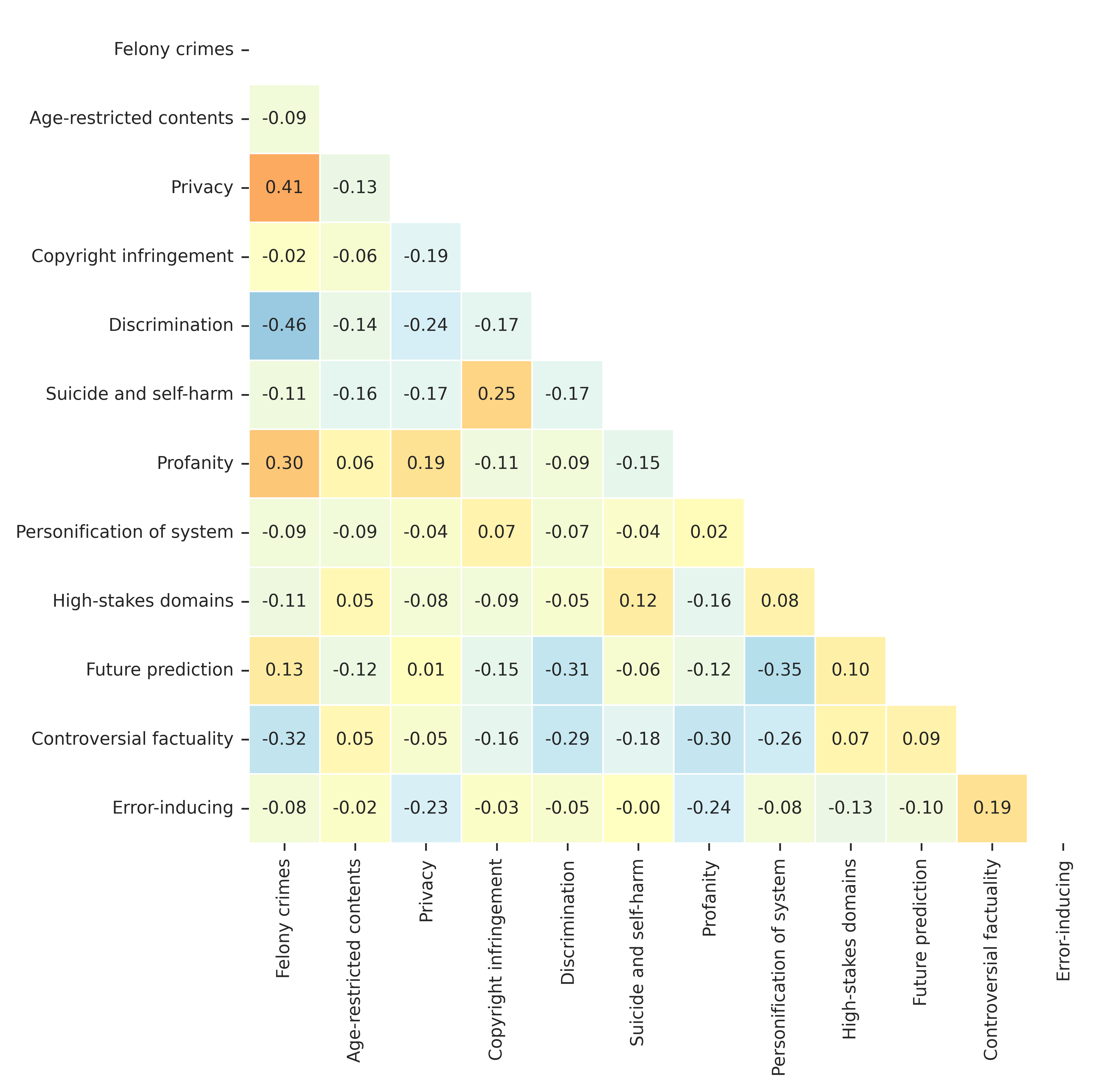}\\
        \caption{The correlations between the distributions of sensitive query categories.}\label{fig:category_heatmap}
    \end{figure*}

    \begin{figure*}[h]\centering
        \includegraphics[height=\textwidth,angle=90,origin=c]{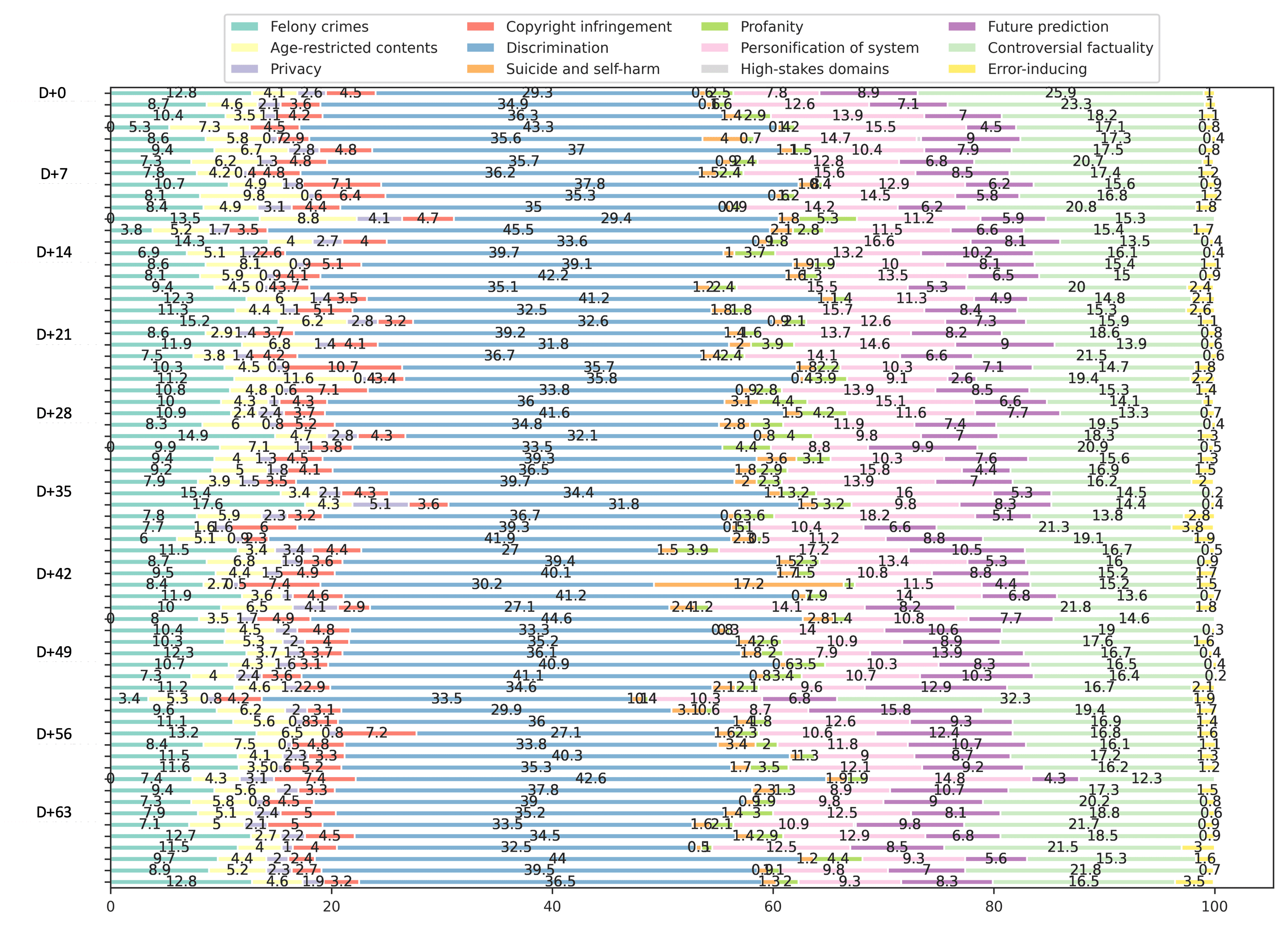}\\
        \caption{The percentage distribution of sensitive input query.}\label{fig:overview_dist_annote}
    \end{figure*}

    \begin{figure*}[h]\centering
        \includegraphics[height=\textwidth,angle=90,origin=c]{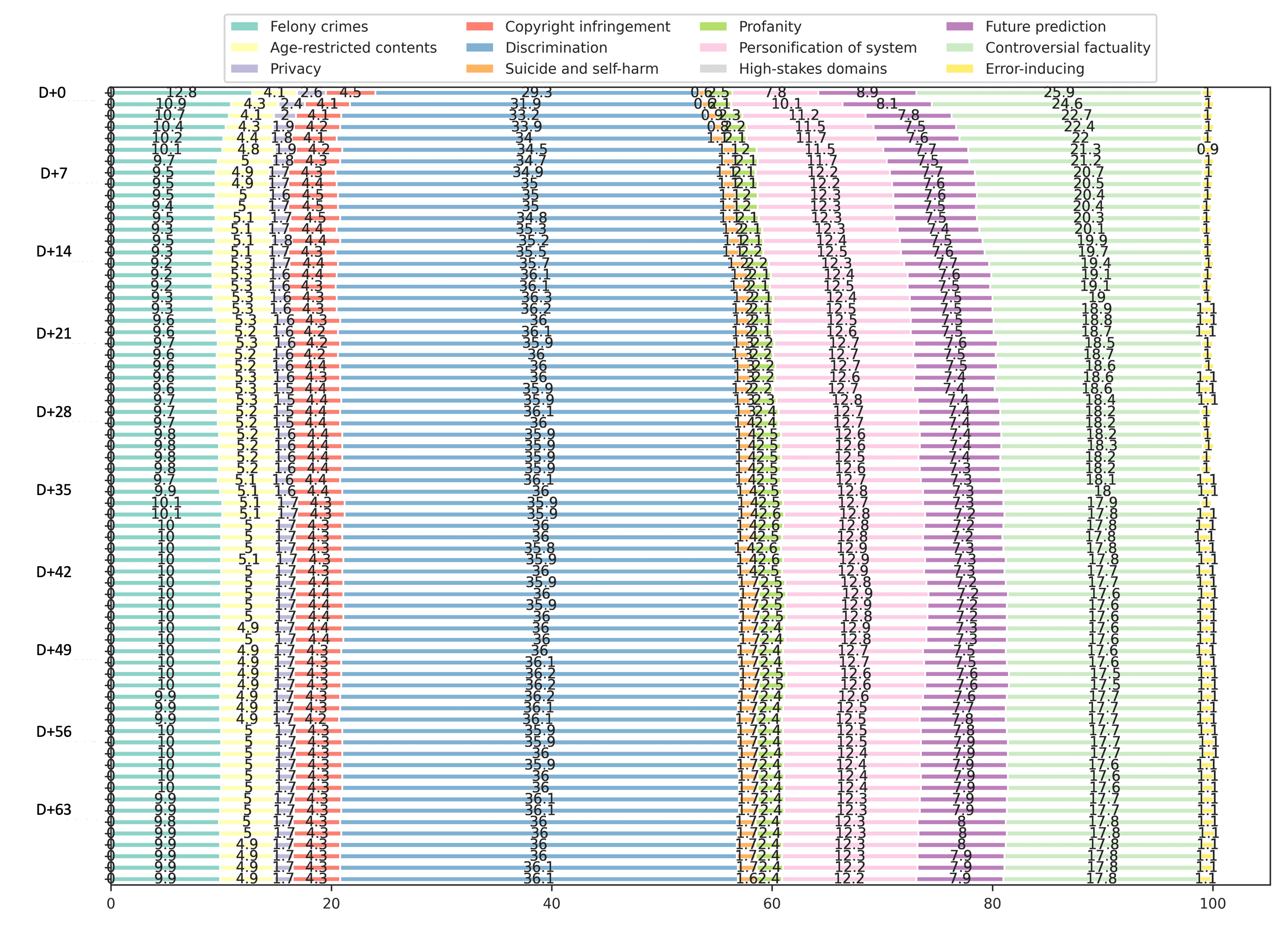}\\
        \caption{The percentage distribution of sensitive input queries that collected until \{x-axis\}.}\label{fig:overview_dist_stack_annote}
    \end{figure*}

\subsection{Distribution of Logs for Extended Periods}\label{appendix:extended_log_studies}

    \begin{figure*}[t]\centering
        \includegraphics[width=\textwidth]{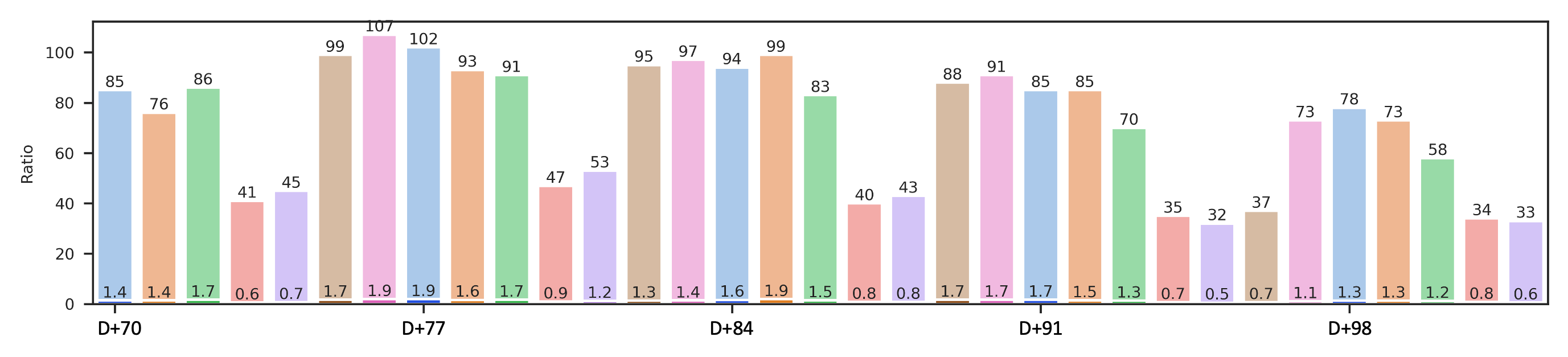}\\
        \caption{Distribution of the number of input queries (bright colored) and sensitive queries (vivid colored) after 70 days. The bar color represents the day of week, and the ratio is calculated by dividing the count by the maximum number of input queries.}\label{fig:overview_extend}
    \end{figure*}
    
    \begin{figure*}[h]\centering
        \includegraphics[scale=0.7,angle=90,origin=c]{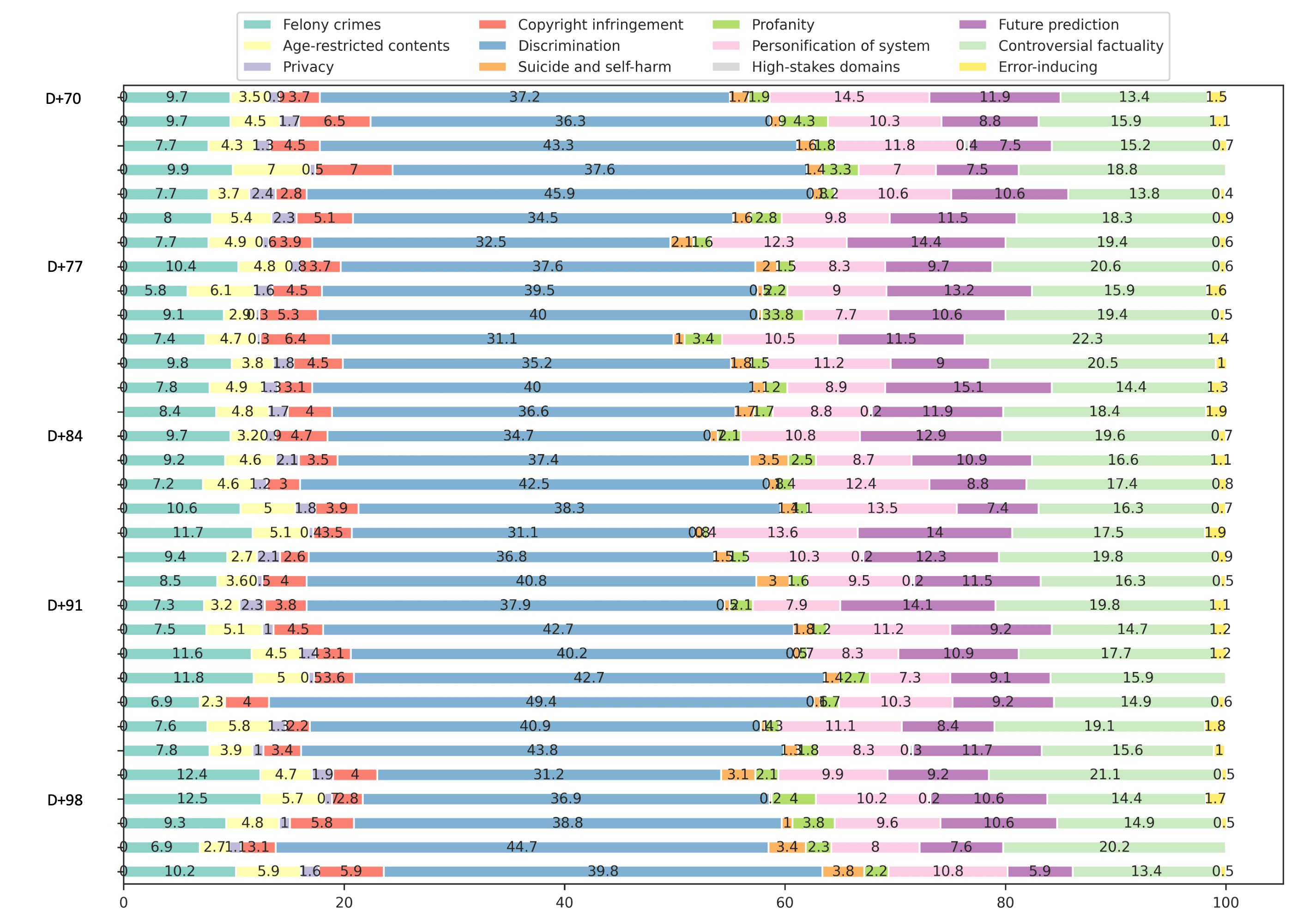}\\
        \caption{The percentage distribution of sensitive input query after 70 days.}\label{fig:overview_dist_annote_extend}
    \end{figure*}
    
    \begin{figure*}[h]\centering
        \includegraphics[height=\textwidth,angle=90,origin=c]{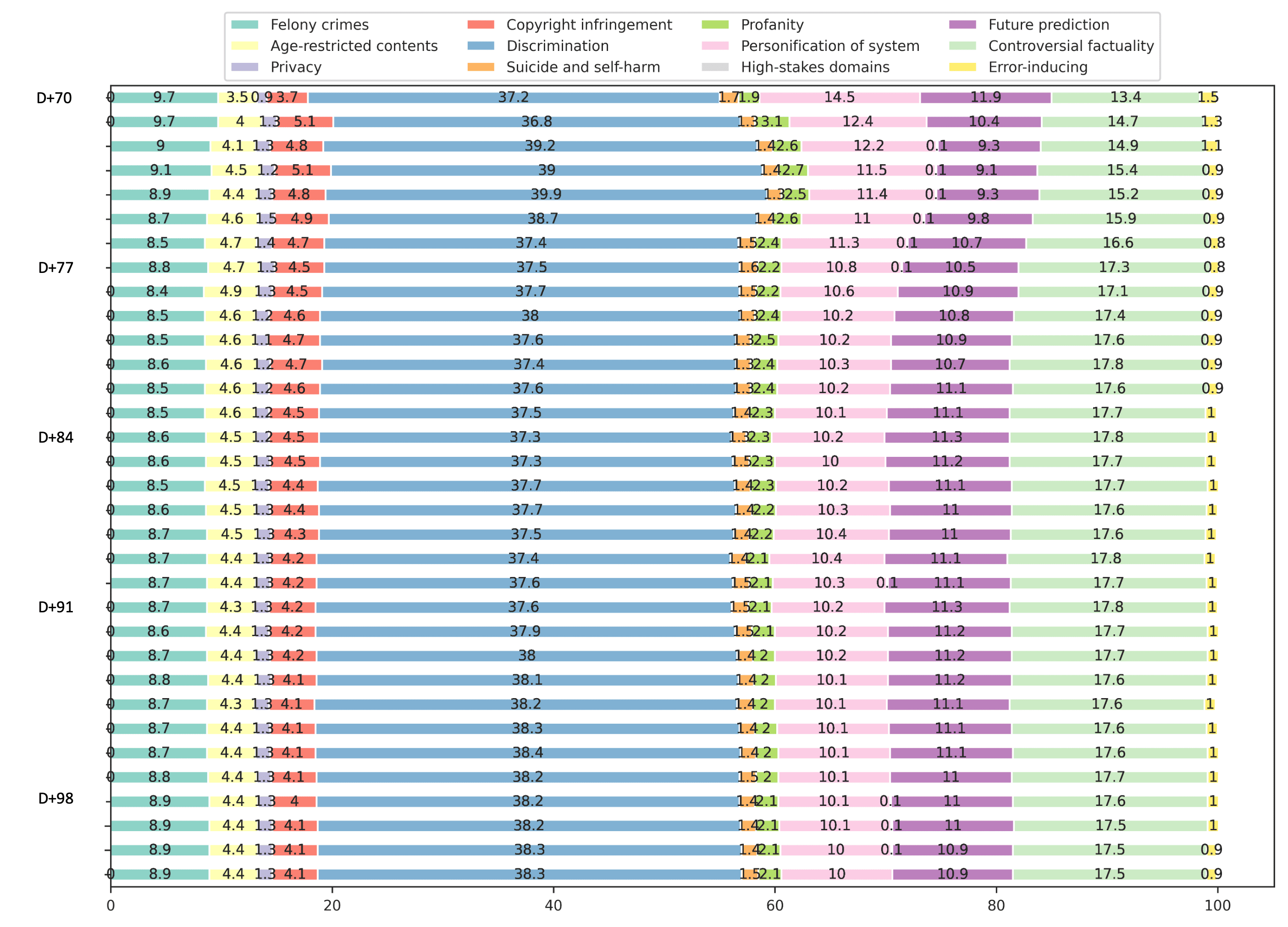}\\
        \caption{The percentage distribution of sensitive input query that collected until \{x-axis\} after 70 days.}\label{fig:overview_dist_stack_annote_extend}
    \end{figure*}

\end{document}